\documentclass[a4paper]{jpconf}
\usepackage{graphicx}
\usepackage[latin1]{inputenc}
\usepackage{amssymb}   
\usepackage{diagbox}
\begin{document}

\hspace{11,8cm} PI/UAN-2017-601FT

\title{Scalar and vector Galileons}

\author{Yeinzon Rodríguez$^{1,2,3}$ and Andrés A. Navarro$^{2}$}

\address{$^1$ Centro de Investigaciones en Ciencias B\'asicas y Aplicadas, Universidad Antonio Nari\~no, \\ Cra 3 Este \# 47A-15, Bogot\'a D.C. 110231, Colombia}
\address{$^2$ Escuela  de  F\'isica,  Universidad  Industrial  de  Santander, \\ Ciudad  Universitaria,  Bucaramanga  680002,  Colombia}
\address{$^3$ Simons Associate at The Abdus Salam International Centre for Theoretical Physics, \\ Strada Costiera 11, I-34151, Trieste, Italy}

\ead{yeinzon.rodriguez@uan.edu.co, navarandres@gmail.com}

\begin{abstract}
An alternative for the construction of fundamental theories is the introduction of Galileons.  These are fields whose action leads to non higher than second-order equations of motion.  As this is a necessary but not sufficient condition to make the Hamiltonian bounded from below, as long as the action is not degenerate, the Galileon construction is a way to avoid pathologies both at the classical and quantum levels.  Galileon actions are, therefore, of great interest in many branches of physics, specially in high energy physics and cosmology.  This proceedings contribution presents the generalities of the construction of both scalar and vector Galileons following two different but complimentary routes.
\end{abstract}

\section{Introduction}
A top-down approach to fundamental physics involves three steps:  1. we have to define the matter content we are interested in, 2. we have to establish what symmetries the action is going to enjoy,  3.  we have to fix things here and there (removing some terms in the action, establishing relations among different coupling constants, etc.) so that no pathologies are present.  This approach is extremely successful with remarkable examples such as the construction of the Standard Model of particle physics, Supersymmetry, and Supergravity.  A complimentary approach consists in formulating the following question:  is there any choice at constructing the fundamental theory?  We can rephrase this question by asking if the removal of any sort of pathologies may lead to a unique action once the matter content and the action symmetries are defined.  When talking about pathologies, we can start with the Ostrogradski's instability \cite{ostro}, that one where the state of the system goes down to lower and lower energy levels because the Hamiltonian is not bounded from below.  Such an instability would lead to disastrous consequences both at the classical and quantum levels \cite{Woodard:2015zca,Woodard:2006nt}.  Ostrogradski's theorem states that, as long as the action is not degenerate, equations of motion higher than second order lead to a Hamiltonian unbounded from below (for a review, see Refs. \cite{Woodard:2015zca,Woodard:2006nt}).  Thus, if we want a healthy fundamental theory, the action has to be constructed so that equations of motion are second-order (at most).  This indeed supports the fact that most of the physics laws are described by second-order differential equations (think, for example, of the Newton laws of mechanics, the Maxwell laws of electromagnetism, and the Einstein equations of gravity).  Of course, this is not a sufficient condition for a healthy theory, so that a Hamiltonian analysis has to be performed once the action has been built.  

In the mid 70's, G. W. Hordenski wrote the most general action for a scalar field and classical gravity that leads to non higher than second-order equations of motion \cite{Horndeski:1974wa}.  A couple of years later, Horndeski himself did the same work exchanging the scalar field for an Abelian vector field \cite{Horndeski:1976gi}.  His results were largely ignored until the first years of the 2010 decade when they were rediscovered \cite{Deffayet:2011gz,Kobayashi:2011nu} in the framework of what is nowadays called Galileons \cite{Nicolis:2008in}.  A Galileon is a scalar field $\pi$ whose action in flat spacetime enjoys a ``Galilean'' symmetry $\pi \rightarrow \pi + b_\mu x^\mu + c$, where $b_\mu$ and $c$ are a constant four-vector and a constant scalar respectively, and whose equation of motion is not higher than second order.  It can be shown that this Galilean symmetry is the responsible for the existence of an equation of motion involving strictly second-order space-time derivatives \cite{Nicolis:2008in}.  A generalization of the Galileon field destroys the Galilean symmetry but allows for lower than second-order space-time derivatives in the equations of motion both in flat and curved spacetime \cite{Deffayet:2013lga,Deffayet:2011gz}.  Such a generalized Galileon is what we will call in this proceedings contribution, from now on, simply as a Galileon.  Since the rediscovery of this idea, its consequences in high energy physics and cosmology have been subject of intensive study.

Life, nonetheless, does not end with scalar fields.  Vector fields are also relevant players in high energy physics and its consequences in cosmology have been under scrutiny in the latest years (see, for example, Refs. \cite{Maleknejad:2012fw,Dimopoulos:2011ws}).  As stated before, Horndeski had studied what we can call now an ``Abelian vector Galileon'':  a vector field whose action is invariant under the $U(1)$ group of symmetries.  In flat spacetime, the only possibility is the Einstein-Hilbert-Maxwell action \cite{Deffayet:2013tca}, but things change in curved spacetime \cite{Horndeski:1976gi}.  As a step towards the fundamental theory, we can relax the gauge-invariance requirement, generalizing this way the Proca action.  It was L. Heisenberg and G. Tasinato, independently, who explored this possibility for the first time \cite{Heisenberg:2014rta,Tasinato:2014eka}.  By following the Galileon construction method described in Refs. \cite{Deffayet:2013lga,Deffayet:2011gz}, these authors built the generalized Proca action by contracting two Levi-Civita tensors with first-order derivatives $\partial_\mu A_\nu$ of the vector field $A_\mu$.  They obtained, this way, a theory where only three degrees of freedom associated to $A_\mu$ propagate.  Moreover, the longitudinal degree of freedom behaves as a scalar Galileon, which means that the theory has a safe decoupling limit.  Such a procedure is powerful but has its limitation:  because of the contraction with the two Levi-Civita tensors, it only produces parity-conserving terms, precluding the construction of allowed parity-violating interactions.  By following a complimentary route, E. Allys, P. Peter and Y. Rodríguez built all the possible Lorentz-invariant terms involving $A_\mu$ and $\partial_\mu A_\nu$ and wrote linear combinations consistent with the propagation of three degrees of freedom for the vector field and a safe decoupling limit \cite{Allys:2015sht}.  The limitation of this procedure is that it does not provide a clear limit in the construction in the sense of a threshold in the number of first-order derivatives of $A_\mu$ in each  term of the action.  In contrast, its advantage is that it does generate parity-violating terms.  Some controversy around these results was initiated with the publication of Ref. \cite{Jimenez:2016isa} but a final agreement was reached in Ref. \cite{Allys:2016jaq}.

The purpose of this paper is to show, schematically, the two Galileon construction procedures described in the previous paragraph, for both a scalar and vector fields.  For the latter, I will show its application to an Abelian vector field and to a set of $SU(2)$ gauge fields.  A short description of the advances in the study of the cosmological implications of the $SU(2)$ vector Galileons will be given at the end.

\section{Scalar Galileon}
Let's assume a scalar field $\pi$ in flat spacetime whose action is Lorentz invariant.  The action may contain the field itself and the first-order and second-order space-time derivatives $\partial_\mu \pi$ and $\partial_\mu \partial_\nu \pi$.  Any higher than second-order space-time derivative would lead, immediately, to higher than second-order equations of motion.  Thus, the action being $S = \int \mathcal{L}(\pi,\partial_\mu \pi,\partial_\mu \partial_\nu \pi) \ d^4x$, the Euler-Lagrange equations become
\begin{equation}
\frac{\partial \mathcal{L}}{\partial \pi} - \partial_\mu \frac{\partial \mathcal{L}}{\partial (\partial_\mu \pi)} + \partial_\mu \partial_\nu \frac{\partial \mathcal{L}}{\partial (\partial_\mu \partial_\nu \pi)} = 0\,, \label{ele}
\end{equation}
which shows that special care has to be taken with those terms in $\mathcal{L}$ involving second-order space-time derivatives.  

The strategy to follow, then, to build the most general action leading to non higher than second-order equations of motion consists in
\begin{enumerate}
\item identifying all the possible Lorentz-invariant terms built from contractions of first-order and second-order space-time derivatives with metric tensors,
\item grouping together all these Lorentz-invariant terms in general linear combinations,
\item establishing relations among the coefficients in the linear combinations so that the higher than second-order contributions produced by the third term in Eq. (\ref{ele}) vanish.
\end{enumerate}
As can be easily checked, the most general action is just given by these linear combinations multiplied by arbitrary functions of $\pi$.

So, what is the structure of these linear combinations?  A careful study of the Euler-Lagrange equations, following the strategy described above, reveals that the adequate combination of Lorentz-invariant terms in the Lagrangian density comes in four different pieces $\mathcal{L}^{\rm Gal}_{N,\pi}$ starting from $N=2$ \cite{Deffayet:2013lga,Deffayet:2011gz}:
\begin{equation}
\mathcal{L}^{\rm Gal}_{N,\pi} \equiv f_N(\pi,X) \; \frac{1}{(4-n)!} \epsilon^{\mu_1 ... \mu_n \sigma_1...\sigma_{4-n}} \epsilon^{\nu_1... \nu_n}_{\;\;\;\;\;\;\;\;\;\;\; \sigma_1... \sigma_{4-n}} \; (\partial_{\mu_1} \partial_{\nu_1} \pi)...(\partial_{\mu_n} \partial_{\nu_n} \pi) \,, \label{master}
\end{equation}
where $N \equiv n+2$, the $\epsilon$ tensors are Levi-Civita, and the $f_N(\pi,X)$ are arbitrary functions of $\pi$ and $X \equiv \partial_\lambda \pi  \partial^\lambda \pi$. Being explicit, the action is written this way:
\begin{equation}
S = \int \sum_{N=2}^5 \mathcal{L}^{\rm Gal}_{N,\pi} \ d^4 x \,,
\end{equation}
where
\begin{eqnarray}
\mathcal{L}^{\rm Gal}_{2,\pi} &\equiv& f_2(\pi,X) \,, \label{2sgal} \\
\mathcal{L}^{\rm Gal}_{3,\pi} &\equiv& f_3(\pi,X) \ \Box \pi \,, \\
\mathcal{L}^{\rm Gal}_{4,\pi} &\equiv& f_4(\pi,X) \ [(\Box \pi)^2 - (\partial_\mu \partial_\nu \pi) (\partial^\mu \partial^\nu \pi)] \,, \\
\mathcal{L}^{\rm Gal}_{5,\pi} &\equiv& f_5(\pi,X) \ [(\Box \pi)^3 - 3(\Box \pi) (\partial_\mu \partial_\nu \pi) (\partial^\mu \partial^\nu \pi) + 2 (\partial_\mu \partial^\nu \pi \ \partial_\nu \partial^\rho \pi \ \partial_\rho \partial^\mu \pi)] \,. \label{5sgal}
\end{eqnarray}

For applications in curved spacetime, the partial derivatives must be replaced by covariant derivatives which, unfortunately, introduce contributions to the equations of motion involving higher than second-order space-time derivatives of both the scalar field and the metric \cite{Deffayet:2009wt,Deffayet:2009mn}.  This issue is easily solved by introducing specific counterterms in $\mathcal{L}^{\rm Gal}_{4,\pi}$ and $\mathcal{L}^{\rm Gal}_{5,\pi}$ \cite{Deffayet:2011gz}.  The action is now written this way:
\begin{equation}
S = \int \left[\sum_{N=2}^5 \mathcal{L}^{\rm Gal}_{N,\pi} + \mathcal{L}^{\rm Gal}_{{\rm Curv},\pi} \right] \ \sqrt{-g}  \ d^4 x \,,
\end{equation}
where
\begin{eqnarray}
\mathcal{L}^{\rm Gal}_{{\rm Curv},\pi} &\equiv& G^{\rm Curv}(\pi) \ G_{\mu \nu} (\nabla^\mu \pi \nabla^\nu \pi) \,, \\
\mathcal{L}^{\rm Gal}_{2,\pi} &\equiv& G_2(\pi,X) \,, \\
\mathcal{L}^{\rm Gal}_{3,\pi} &\equiv& G_3(\pi,X) \ \Box \pi \,, \\
\mathcal{L}^{\rm Gal}_{4,\pi} &\equiv& G_4(\pi,X) R + G_{4,X} \ [(\Box \pi)^2 - (\nabla_\mu \nabla_\nu \pi) (\nabla^\mu \nabla^\nu \pi)] \,, \\
\mathcal{L}^{\rm Gal}_{5,\pi} &\equiv& G_5(\pi,X) G_{\mu \nu} (\nabla^\mu \nabla^\nu \pi) \nonumber \\
&& - \frac{1}{6} G_{5,X} \ [(\Box \pi)^3 - 3(\Box \pi) (\nabla_\mu \nabla_\nu \pi) (\nabla^\mu \nabla^\nu \pi) + 2 (\nabla_\mu \nabla^\nu \pi \ \nabla_\nu \nabla^\rho \pi \ \nabla_\rho \nabla^\mu \pi)] \,, \nonumber \\
&&
\end{eqnarray}
where $g$ is the determinant of the metric, $R$ is the Ricci scalar, $G_{\mu \nu}$ is the Einstein tensor, $\nabla$ stands for a covariant derivative, and  $G_{N,X} \equiv \partial G_N / \partial X$.  As we can see, a new term is introduced, $\mathcal{L}^{\rm Gal}_{{\rm Curv},\pi}$, which vanish in flat spacetime;  since $G_{\mu \nu}$ is divergenceless, this new term, likewise the counterterms in $\mathcal{L}^{\rm Gal}_{4,\pi}$ and $\mathcal{L}^{\rm Gal}_{5,\pi}$, does not introduce new propagating degrees of freedom in the tensor sector \cite{Horndeski:1976gi,Jimenez:2013qsa}.

It is clear that Eq. (\ref{master}) is just a consequence of examining the Euler-Lagrange equations in the search for non higher than second-order equations of motion.  Following the strategy described above or employing directly Eq. (\ref{master}) are, therefore, equivalent procedures to build the action for a scalar Galileon.  Such an equivalence is not necessarily present, however, when constructing vector Galileons, as the next section shows.

\section{Vector Galileons}

\subsection{The generalized Abelian Proca action}
Let's start in flat spacetime.  Because of the Helmholtz decomposition, any space-time vector $A_\mu$ can be split in two pieces:
\begin{equation}
A_\mu = \mathcal{A}_\mu + \partial_\mu \pi \,,
\end{equation}
where $\mathcal{A}_\mu$ is divergenceless and $\pi$ is a scalar field.  $\pi$ is, therefore, the longitudinal degree of freedom associated to $A_\mu$.  Thus, should $A_\mu$ be identified as a vector Galileon, $\pi$ must be a scalar Galileon.  The immediate consequence of this fact is that the action of the vector field may include the field itself and its first-order space-time derivative $\partial_\mu A_\nu$ only.  This is welcome since the equation of motion for $A_\mu$ will not be higher than second order no matter the specific form of the action.

The first route to build the Galilean action follows similar lines to those described for the scalar Galileon case:
\begin{enumerate}
\item identify all the possible Lorentz-invariant terms built from contractions of vector fields and first-order space-time derivatives with the primitive invariants of the Lorentz group ($SO(3,1)$):  metric tensors and, at most, one Levi-Civita tensor,
\item group together all these Lorentz-invariant terms in general linear combinations,
\item establish relations among the coefficients in the linear combinations so that no more than three degrees of freedom propagate,
\item make the replacement $A_\mu \rightarrow \partial_\mu \pi$ and remove all those terms whose resultant action does not correspond to that of a scalar Galileon.
\end{enumerate}
The restriction to the number of Levi-Civita tensors in the first step comes from the fact that the product of two such tensors can always be expressed as a linear combination of products of metric tensors.  Notice that, because of the antisymmetric properties of the Levi-Civita tensor, the analogous step for a scalar Galileon only requires contractions with metric tensors.  Regarding the third step, the propagation of just three degrees of freedom, they being the spatial components of the vector field, is established by the Hessian condition $\mathcal{H}^{0 \nu} = 0$ \cite{Heisenberg:2014rta} where
\begin{equation}
\mathcal{H}^{\mu \nu} \equiv \frac{\partial^2 \mathcal{L}}{\partial(\partial_0 A_\mu) \partial(\partial_0 A_\nu)} \,.
\end{equation}
Thus, the action for the vector Galileon, which generalizes the Abelian Proca action, turns out to be
\begin{equation}
S = \int \left[-\frac{1}{4} F_{\mu \nu} F^{\mu \nu} + \frac{1}{2} m^2 A^2 + \sum_{N=2}^6 \mathcal{L}^{\rm Gal}_{N,A} \right] \ d^4 x \,, \label{actionA}
\end{equation}
where \cite{Jimenez:2016isa,Allys:2016jaq}
\begin{eqnarray}
\mathcal{L}^{\rm Gal}_{2,A} &\equiv& f_2(A_\mu,F_{\mu \nu},\tilde{F}_{\mu \nu}) = f_2 \left[A^2,F^2,F\cdot \tilde{F},(A\cdot \tilde{F})^2\right] \,, \label{L2A} \\
\mathcal{L}^{\rm Gal}_{3,A} &\equiv& f_3(A^2) \ S^\mu_{\;\; \mu} \,, \\
\mathcal{L}^{\rm Gal}_{4,A} &\equiv& f_4(A^2) \ [(S^\mu_{\;\; \mu})^2 - S_\rho^{\;\; \sigma} S_\sigma^{\;\; \rho}] \,, \\
\mathcal{L}^{\rm Gal}_{5,A} &\equiv& f_5(A^2) \ [(S^\mu_{\;\; \mu})^3 - 3 (S^\mu_{\;\; \mu}) S_\rho^{\;\; \sigma} S_\sigma^{\;\; \rho} + 2 S_\rho^{\;\; \sigma} S_\sigma^{\;\; \gamma} S_\gamma^{\;\; \rho} ] + g_5(A^2) \ \tilde{F}^{\alpha \mu} \tilde{F}^\beta_{\;\; \mu} S_{\alpha \beta} \,, \\
\mathcal{L}^{\rm Gal}_{6,A} &\equiv& g_6(A^2) \ \tilde{F}^{\alpha \beta} \tilde{F}^{\mu \nu} S_{\alpha \mu} S_{\beta \nu} \,. \label{L6A}
\end{eqnarray}
In contrast to the local gauge field theory, a new element appears in the action accompanying $A_\mu$, the field strength tensor $F_{\mu \nu} \equiv \partial_\mu A_\nu - \partial_\nu A_\mu$, and its Hodge dual $\tilde{F}_{\mu \nu} \equiv \frac{1}{2} \epsilon_{\mu \nu \rho \sigma} F^{\rho \sigma}$.  This new element is the symmetric version of the field strength tensor:  $S_{\mu \nu} \equiv \partial_\mu A_\nu + \partial_\nu A_\mu$.  It is important to notice that the Lagrangian pieces whose coefficient is a function $f_N$ reduce to the respective scalar Galileon Lagrangians of Eqs. (\ref{2sgal}) - (\ref{5sgal}) when $A_\mu$ is replaced by $\partial_\mu \pi$ (except for the possible $\pi$ dependence of $f_N$).  Meanwhile, those Lagrangian pieces premultiplied by a function $g_N$ vanish when we make the same replacement;  this is why the generalized Proca action has an additional Lagrangian, $\mathcal{L}^{\rm Gal}_{6,A}$, compared to its scalar counterpart, which was not discovered in the pioneering papers of L. Heisenberg and G. Tasinato \cite{Heisenberg:2014rta,Tasinato:2014eka}.

This way of procedure can be extended for an arbitrary number of first-order derivatives without a complete proof why the sequence should stop.  This indeed led to E. Allys, P. Peter, and Y. Rodríguez, in Ref. \cite{Allys:2015sht}, to construct all the Lagrangian pieces up to $\mathcal{L}^{\rm Gal}_{7,A}$ and to conjecture that an infinite tower of terms would be generated;  however, as shown in Ref. \cite{Jimenez:2016isa}, the $\mathcal{L}^{\rm Gal}_{7,A}$ Lagrangian presented in Ref. \cite{Allys:2015sht} identically vanishes.  This is an unfortunate limitation of the procedure already described which is not present in the second route we will present shortly.  The advantage of the method already described is that it generates parity-violating terms:  since some of the Lorentz-invariant terms in the first step are built from contractions with just one Levi-Civita tensor, parity-violating terms are inevitable.  Most of these terms end up encoded in $\mathcal{L}^{\rm Gal}_{2,A}$ while the others vanish identically (at least, up to $\mathcal{L}^{\rm Gal}_{7,A}$) \cite{Allys:2016jaq}.  There exists a special parity-violating term uncovered in Ref. \cite{Allys:2016jaq} which is impossible to build following the second route.  This term is specially different to the other terms in the Lagrangian because not all the $A_\mu$ pairs come as $A^2$:
\begin{equation}
\mathcal{L}^{\rm Gal,bis}_{4,A} \equiv g_4(A^2) \ A^\mu \tilde{F}_{\mu \nu} S^{\nu \lambda} A_\lambda \,.
\end{equation}
This term, however, is redundant (it already belongs to $\mathcal{L}^{\rm Gal}_{2,A}$) since $\tilde{F}_{\mu \nu}$ is divergenceless (see the Appendix).

The curved spacetime version of the generalized Proca action is obtained, as in the scalar Galileon case, by replacing the standard space-time derivatives by covariant derivatives and adding the required counterterms.  The latter are chosen in order to avoid higher than second-order equations of motion and the wrong number of propagating degrees of freedom at every level of the decoupling limit.  Refs. \cite{Heisenberg:2014rta,Allys:2015sht,Jimenez:2016isa} arrive to the following result:
\begin{equation}
S = \int \left[-\frac{1}{4} F_{\mu \nu} F^{\mu \nu} + \frac{1}{2} m^2 A^2 + \sum_{N=2}^6 \mathcal{L}^{\rm Gal}_{N,A} + \mathcal{L}^{\rm Gal}_{{\rm Curv},A} \right] \ \sqrt{-g}\ d^4 x \,,
\end{equation}
where
\begin{eqnarray}
\mathcal{L}^{\rm Gal}_{{\rm Curv},A} &\equiv& f^{\rm Curv} G_{\mu \nu} A^\mu A^\nu \,, \\
\mathcal{L}^{\rm Gal}_{2,A} &\equiv& f_2(A_\mu,F_{\mu \nu},\tilde{F}_{\mu \nu}) = f_2 \left[A^2,F^2,F\cdot \tilde{F},(A\cdot \tilde{F})^2\right] \,, \\
\mathcal{L}^{\rm Gal}_{3,A} &\equiv& f_3(A^2) \ S^\mu_{\;\; \mu} \,, \\
\mathcal{L}^{\rm Gal}_{4,A} &\equiv& f_4(A^2) R + \frac{1}{2} f_{4,A^2} \ [(S^\mu_{\;\; \mu})^2 - S_\rho^{\;\; \sigma} S_\sigma^{\;\; \rho}] \,, \\
\mathcal{L}^{\rm Gal}_{5,A} &\equiv& f_5(A^2) G_{\mu \nu} S^{\mu \nu} + \frac{3}{4} f_{5,A^2} \ [(S^\mu_{\;\; \mu})^3 - 3 (S^\mu_{\;\; \mu}) S_\rho^{\;\; \sigma} S_\sigma^{\;\; \rho} + 2 S_\rho^{\;\; \sigma} S_\sigma^{\;\; \gamma} S_\gamma^{\;\; \rho} ]  \nonumber \\
&& + g_5(A^2) \ \tilde{F}^{\alpha \mu} \tilde{F}^\beta_{\;\; \mu} S_{\alpha \beta} \,, \\
\mathcal{L}^{\rm Gal}_{6,A} &\equiv& g_6(A^2) L_{\mu \nu \rho \sigma} F^{\mu \nu} F^{\rho \sigma} + g_{6,A^2} \ \tilde{F}^{\alpha \beta} \tilde{F}^{\mu \nu} S_{\alpha \mu} S_{\beta \nu} \,,
\end{eqnarray}
where $f^{\rm Curv}$ is just a constant scalar and $L_{\mu \nu \rho \sigma}$ is the double dual Riemann tensor:
\begin{equation}
L^{\mu \nu \alpha \beta} \equiv \frac{1}{4} \epsilon^{\mu \nu \rho \sigma} \epsilon^{\alpha \beta \gamma \delta} R_{\rho \sigma \gamma \delta} \,,
\end{equation}
$R_{\rho \sigma \gamma \delta}$ being the Riemann tensor.

The second route to build the generalized Proca action consists in extending Eq. (\ref{master}), valid for a scalar Galileon, to the vector field $A_\mu$.  This means contracting products of first-order space-time derivatives of $A_\mu$ with two Levi-Civita tensors:
\begin{equation}
\mathcal{L}^{\rm Gal}_{N,A} \propto \epsilon^-{}_- \epsilon^-{}_- \
 \partial_\centerdot A_\centerdot \partial_\centerdot A_\centerdot\cdots \,. \label{master2}
\end{equation}
What is interesting about this construction is that, as shown in Ref. \cite{Allys:2016jaq}, the Hessian condition is satisfied automatically.  In addition, when making the replacement $A_\mu \rightarrow \partial_\mu \pi$, Eq. (\ref{master2}) reduces to Eq. (\ref{master}), except for $N = 6$, so that the theory has a safe decoupling limit (for $N = 6$, we have to discard that term that does not vanish when taking the scalar limit, see Ref. \cite{Jimenez:2016isa}).  Eq. (\ref{master2}) reproduces Eqs. (\ref{L2A}) - (\ref{L6A}) except for the fact that Eq. (\ref{L2A}) has to be understood, in this approach, as containing only parity-conserving terms.  This is the limitation of this procedure:  since the Lagrangian pieces are built from contractions with two Levi-Civita tensors, it is impossible to build parity-violating terms.  There is however an even bigger limitation:  there does not exist a proof that Eq. (\ref{master2}) generates all the possible Lagrangian pieces of the generalized Proca action in contrast with Eq. (\ref{master}) for which there does exist a formal proof (see Ref. \cite{Deffayet:2011gz}).  Since all the contractions of space-time derivatives of $A_\mu$ with the primitive invariants of the Lorentz groups can be written as contractions of the former with an unrestricted number of Levi-Civita tensors, there is no reason why the number of Levi-Civita tensors should be only two.  Thus, although the second route gives a finite number of Lagrangian pieces, since the number of space-time indices in the Levi-Civita tensors contracted with space-time derivatives is saturated in $\mathcal{L}^{\rm Gal}_{6,A}$, it might happen that the actual number of Lagrangian pieces contributing to the generalized Proca action is larger.
 
\subsection{The generalized $SU(2)$ Proca action} 
Let's now consider a set of three vector fields $A_\mu^a$, with $a$ running from 1 to 3, whose action is invariant under the $SU(2)$ global symmetry group.  Grouping these vector fields into a single matrix $\mathcal{A}_\mu = A_\mu^a T_a$, where the $T_a$ are the matrix generators of the $SU(2)$ transformations, it is possible to show that $\mathcal{A}_\mu$ transforms in the respective adjoint representation (see, for example, Ref. \cite{Rodolfo}):
\begin{equation}
\mathcal{A}_\mu' = e^{ig \vec{\epsilon} \cdot \vec{T}} \mathcal{A}_\mu e^{-ig \vec{\epsilon} \cdot \vec{T}} \,.
\end{equation}
In the previous expression, $g$ is the coupling constant, $\vec{\epsilon}$ is a three-dimensional vector that parametrizes the amount of the transformation, and $\vec{T}$ is the ``vector'' built with the matrix generators.  Such generators satisfy the Lie algebra
\begin{equation}
[T_a,T_b] = \epsilon_{ab}^{\;\;\;\; c} \ T_c \,,
\end{equation}
where $\epsilon_{abc}$ denotes the structure constants of the group which, for $SU(2)$, correspond to the Levi-Civita symbol.

With these preliminaries in mind, the first route to build the generalized $SU(2)$ Proca action consists in the following steps:
\begin{enumerate}
\item identify all the possible Lorentz-invariant terms built from contractions of vector fields $A_\mu$ (without $SU(2)$ group indices) and first-order space-time derivatives $\partial_\mu A_\nu$ with the primitive invariants of the Lorentz group ($SO(3,1)$):  metric tensors and, at most, one Levi-Civita tensor,
\item add $SU(2)$ group indices to the terms identified in the previous step and contract with the primitive invariants of the $SU(2)$ group:  group induced metrics $g_{ab}$ and structure constants $\epsilon_{abc}$ (at most, one Levi-Civita symbol).
\item group together all these Lorentz-$SU(2)$ invariant terms in general linear combinations,
\item establish relations among the coefficients in the linear combinations so that no more than three degrees of freedom propagate,
\item make the replacement $A_\mu^a \rightarrow \partial_\mu \pi^a$ and remove all those terms whose resultant action does not correspond to that of multi-scalar Galileons in the three-dimensional representation of $SU(2)$ (see Ref. \cite{Allys:2016hfl} and Appendix A of Ref. \cite{Allys:2016kbq};  see also Ref. \cite{Padilla:2010ir}).
\end{enumerate}
Because of symmetry reasons, some of the terms found in the first step vanish identically;  however, they must be kept since the addition of group indices in the second step can make them non vanishing.
The restriction to the number of Levi-Civita symbols in the second step comes from the fact that the product of two such symbols can always be expressed as a linear combination of products of group induced metrics \cite{Metha:1983mng}.

This procedure was followed in Ref. \cite{Allys:2016kbq} with the imposed restriction of, at most, six space-time indices in the Lorentz-$SU(2)$ invariant terms.  Such a restriction is just technical in its nature:  it was imposed because the number of Lorentz-$SU(2)$ invariant terms scales strongly when allowing for more indices, as Table \ref{neverending} shows.
\begin{table}
\caption{\label{neverending} Number of Lorentz-$SU(2)$ invariant terms built from contractions of some number of vector fields (first line) and some number of first-order derivatives (first column) with the primitive invariants of $SO(3,1)$ and $SU(2)$.}
\begin{center}
\begin{tabular}{|c|c|c|c|}
\hline
\backslashbox{$\# \partial^\mu A^{\nu a}$}{$\# A^{\rho b}$} & 0 & 2 & 4
\\
\hline
1 & 0  & 3 & 36  \\
\hline
2 & 4  & 42  & 510 \\
\hline
3 & 9  & 312  &  \\
\hline
\end{tabular}
\end{center}
\end{table}	

Thus, the flat spacetime generalized $SU(2)$ Proca action is given by
\begin{equation}
S = \int \left[-\frac{1}{4} F_{\mu \nu}^a F^{\mu \nu}_a + \frac{1}{2} m^2 A^\mu_a A_\mu^a + \sum_{N=2}^4 \mathcal{L}^{\rm Gal}_{N,A} \right] \ d^4 x \,,
\end{equation}
where \cite{Allys:2016kbq}
\begin{eqnarray}
\mathcal{L}^{\rm Gal}_{2,A} &\equiv& f_2(A_\mu^a,F_{\mu \nu}^a,\tilde{F}_{\mu \nu}^a) \,,  \\
\mathcal{L}^{\rm Gal}_{3,A} &\equiv& 0 \,, \\
\mathcal{L}^{\rm Gal}_{4,A} &\equiv& f_4^1 \ \Big\{ (A_b \cdot A^b) \left[ \left(\partial \cdot A_a \right)\left(\partial \cdot A^a \right) - (\partial_\mu A^\nu_a)(\partial^\mu A_\nu^a) \right]  \nonumber \\
&& + 2 (A_a \cdot A_b) [\left(\partial \cdot A^a \right) (\partial \cdot A^b) - (\partial_\mu A^{\nu a})(\partial^\mu A_\nu^b)] \Big\} \nonumber \\
&& + f_4^2 \ \Big\{ (A_a \cdot A_b) [\left(\partial \cdot A^a \right) (\partial \cdot A^b) - (\partial_\mu A^{\nu a})(\partial^\mu A_\nu^b)] \nonumber \\
&& + (A^{\mu a} A^{\nu b}) \left[\left(\partial_\mu A^\alpha_a \right)\left(\partial_\nu A_{\alpha b} \right) - \left(\partial_\nu A^\alpha_a \right)\left(\partial_\mu A_{\alpha b} \right) \right] \Big\} \nonumber \\
&& + f_4^3 \ \tilde{G}_{\mu\sigma}^b A^\mu_a A_{\alpha b}  S^{\alpha\sigma a} \,, \label{L4naA}
\end{eqnarray}
where the $f_4$ are arbitrary constant scalars, $F_{\mu \nu}^a$ is the non-Abelian field strength tensor:
\begin{equation}
F_{\mu \nu}^a \equiv \partial_\mu A_\nu^a  - \partial_\nu A_\nu^a + g \epsilon^a{}_{b c} A_\mu^b A_\nu ^c \,,
\end{equation}
$G_{\mu \nu}^a$ is the Abelian version of $F_{\mu \nu}^a$:
\begin{equation}
G_{\mu \nu}^a \equiv \partial_\mu A_\nu^a  - \partial_\nu A_\nu^a \,,
\end{equation}
and $S_{\mu \nu}^a$ is the symmetrized version of $G_{\mu \nu}^a$:
\begin{equation}
S_{\mu \nu}^a \equiv \partial_\mu A_\nu^a  + \partial_\nu A_\nu^a \,.
\end{equation}
Although $\mathcal{L}^{\rm Gal}_{5,A}$ and $\mathcal{L}^{\rm Gal}_{6,A}$ were not built, a comparison with the Abelian case allowed the authors of Ref. \cite{Allys:2016kbq} to conjecture that
\begin{equation}
\mathcal{L}^{\rm Gal}_{5,A} \equiv f_5 \ \epsilon_{a b c} \left(A^a \cdot A^d\right) \tilde{G}^{\alpha\mu}_d\tilde{G}^{\beta}{}_\mu^b S_{\alpha\beta}^c \,, \label{conjecture5A}
\end{equation}
and
\begin{equation}
\mathcal{L}_{6,A}^{\rm Gal} \equiv f_6^1 \ \tilde{G}^{\alpha\beta}_a \tilde{G}^{\mu\nu a} S_{\alpha\mu}^b S_{\beta\nu b} + f_6^2 \ \tilde{G}^{\alpha\beta}_a \tilde{G}^{\mu\nu}_b S_{\alpha\mu}^a S_{\beta\nu}^b \,,  \label{conjecture6A}
\end{equation}
where $f_5$ and the $f_6$ are arbitrary constant scalars.
It is very interesting to notice that the non-Abelian version of $\mathcal{L}^{\rm Gal,bis}_{4,A}$, the last term in Eq. (\ref{L4naA}), is not a redundant term because the group indices protect it;  this is, then, the first parity-violating term that is not included in $f_2$.

The curved spacetime version of this action is a bit more elaborate in its counterterms:
\begin{equation}
S = \int \left[-\frac{1}{4} F_{\mu \nu}^a F^{\mu \nu}_a + \frac{1}{2} m^2 A^\mu_a A_\mu^a + \sum_{N=2}^4 \mathcal{L}^{\rm Gal}_{N,A} + \sum_{m = 1}^5 \mathcal{L}^{\rm Gal}_{{\rm Curv},m,A} \right] \sqrt{-g} \ d^4 x \,,  \label{greatmaster}
\end{equation}
where \cite{Allys:2016kbq}
\begin{eqnarray}
\mathcal{L}^{\rm Gal}_{{\rm Curv},1,A} &\equiv& f^{\rm Curv}_1 G_{\mu \nu} A^{\mu a} A^\nu_a \,, \\
\mathcal{L}^{\rm Gal}_{{\rm Curv},2,A} &\equiv& f^{\rm Curv}_2  L_{\mu \nu \rho \sigma} F^{\mu \nu a} F^{\rho \sigma}_a \,, \label{rusos} \\
\mathcal{L}^{\rm Gal}_{{\rm Curv},3,A} &\equiv& f^{\rm Curv}_3  \epsilon_{a b c} L_{\mu \nu \rho \sigma} F^{\mu \nu a} A^{\rho b} A^{\sigma c} \,, \\
\mathcal{L}^{\rm Gal}_{{\rm Curv},4,A} &\equiv& f^{\rm Curv}_4 L_{\mu \nu \rho \sigma} A^{\mu a} A^\nu_a A^{\rho b} A^\sigma_b \,, \\
\mathcal{L}^{\rm Gal}_{2,A} &\equiv& f_2(A_\mu^a,F_{\mu \nu}^a,\tilde{F}_{\mu \nu}^a) \,,  \\
\mathcal{L}^{\rm Gal}_{3,A} &\equiv& 0 \,, \\
\mathcal{L}^{\rm Gal}_{4,A} &\equiv& f_4^1 \ \Big\{ (A_b \cdot A^b) \left[ \left(\nabla\cdot A_a \right)\left(\nabla\cdot A^a \right) - (\nabla_\mu A^\nu_a)(\nabla^\mu A_\nu^a) + \frac{1}{4} (A_a \cdot A^a) R \right]  \nonumber \\
&& + 2 (A_a \cdot A_b) \left[\left(\nabla\cdot A^a \right) (\nabla\cdot A^b) - (\nabla_\mu A^{\nu a})(\nabla^\mu A_\nu^b) + \frac{1}{2} (A^a \cdot A^b) R \right] \Big\} \nonumber \\
&& + f_4^2 \ \Big\{ (A_a \cdot A_b) \left[\left(\nabla\cdot A^a \right) (\nabla\cdot A^b) - (\nabla_\mu A^{\nu a})(\nabla^\mu A_\nu^b) + \frac{1}{4} (A^a \cdot A^b) R \right] \nonumber \\
&& + (A^{\mu a} A^{\nu b}) \left[\left(\nabla_\mu A^\alpha_a \right)\left(\nabla_\nu A_{\alpha b} \right) - \left(\nabla_\nu A^\alpha_a \right)\left(\nabla_\mu A_{\alpha b} \right) - \frac{1}{2} A^\rho_a A^\sigma_b R_{\mu \nu \rho \sigma} \right] \Big\} \nonumber \\
&& + f_4^3 \ \tilde{G}_{\mu\sigma}^b A^\mu_a A_{\alpha b}  S^{\alpha\sigma a} \,, \label{nonabpvt}
\end{eqnarray}
where the $f^{\rm Curv}$ are just constant scalars.

The second route to build the generalized $SU(2)$ Proca action is employed in Ref. \cite{Jimenez:2016upj}.  Two Levi-Civita tensors are contracted with products of vector fields and their first-order derivatives:
\begin{equation}
\mathcal{L} \propto \epsilon^{\mu \nu \alpha \beta} \epsilon^{\rho \sigma \gamma \delta} \partial_\mu A_\rho^a \cdots A_\nu^b A_\sigma^c \cdots \,,  \label{lnab}
\end{equation}
the free Lorentz indices being appropriately contracted with space-time metrics and the group indices being contracted with the primitive invariants of the $SU(2)$ group.  Similarly to the Abelian case, this construction automatically satisfies the Hessian condition $\mathcal{H}^{0 \nu a b} = 0$ where  \cite{Allys:2016kbq} 
\begin{equation}
\mathcal{H}^{\mu \nu a b} \equiv \frac{\partial^2 \mathcal{L}}{\partial(\partial_0 A_{\mu a}) \partial(\partial_0 A_{\nu b})} \,,
\end{equation}
so that just three degrees of freedom are able to propagate.  In adition, the scalar limit has to be taken and checked in order to be sure that the theory has a safe decoupling limit.  Unfortunately, Ref.  \cite{Jimenez:2016upj} lacks for this last step.  The same pros and contras of this route discussed in previous pages apply here as well.  We stress that there does not exist a formal proof that applying Eq. (\ref{lnab}) is equivalent to the first route described above;  indeed, such a proof cannot exist due to the existence of the non redundant parity-violating term in $\mathcal{L}^{\rm Gal}_{4,A}$ (the last one in Eq. (\ref{nonabpvt})).

\subsubsection{Cosmological implications}
We will now talk about some advances already made in the exploration of the cosmological implications of the action in Eq. (\ref{greatmaster}) \cite{andres}.  The idea is to have a ``cosmic triad'' configuration, i.e., the three vector fields orthogonal to each other and of the same norm, with the temporal components vanishing, in a Friedmann-Lemaitre-Robertson-Walker background.  Such a configuration avoids any kind of anisotropy, both at the background and the perturbative level \cite{Rodriguez:2015xra}, which are severely restricted by the observations \cite{Kim:2013gka,Ramazanov:2013wea,Ade:2015hxq,Ramazanov:2016gjl}.  It has also been studied previously with success in the Gauge-flation \cite{Maleknejad:2011jw,Maleknejad:2011sq,Nieto:2016gnp} and Chromo-natural inflation \cite{Adshead:2012kp} models.  
We have discovered that all the Lagrangian pieces in Eq. (\ref{greatmaster}) admit this configuration, except for the parity-violating term in the last line of Eq. (\ref{nonabpvt}) and, in principle, some terms belonging to $\mathcal{L}^{\rm Gal}_{2,A}$.  For its own existence, the parity-violating term requires not only at least one non vanishing temporal component but also a non orthogonal configuration.  The Lagrangian in Eq. (\ref{conjecture5A}) shares the same requirements.  In contrast, the Lagrangian in Eq. (\ref{conjecture6A}) admits the orthogonality but requires the existence of the temporal components;  this implies momentum flow in the energy-momentum tensor which destroys the required isotropy.  The Lagrangian in Eq. (\ref{rusos}) had already been studied in Ref. \cite{Davydov:2015epx}, finding that a short period of inflation is generated;  however, as shown in Ref. \cite{BeltranJimenez:2017cbn}, this Lagrangian is plagued by ghosts and Laplacian instabilities.  The other $\mathcal{L}^{\rm Gal}_{\rm Curv, A}$, except for $\mathcal{L}^{\rm Gal}_{{\rm Curv},1,A}$ lead to very similar results at the background level and, probably, will have the same fate as $\mathcal{L}^{\rm Gal}_{{\rm Curv},2,A}$ at the perturbative level.  Regarding $\mathcal{L}^{\rm Gal}_{{\rm Curv},1,A}$, it does not lead to any interesting cosmology at the background level.  We are still trying to uncover the cosmological implications of the Lagrangian in Eq. (\ref{nonabpvt}) (except for the parity-violating term).

\section{Conclusions}
Two different but complimentary routes were exposed that lead to the construction of scalar and vector Galileon actions.  Both routes are equivalent when building the Galileon action for a scalar field;  however, such equivalence does not necessarily exist when building vector Galileons.  Although the first route produces both parity-conserving and parity-violating terms, it does not give any clear end to the sequence of Lagrangian pieces that can be built.  Indeed, it is possible, although unlikely, that such a sequence goes to infinity.  In contrast, the second route does not produce parity-violating terms but it does provide a clear limit for the sequence of Lagrangians.  Unfortunately, the second route is based on an unproved assumption.  This does not preclude, anyway, the enormous help that the second route gives us to build and understand the structure of the Galileon actions.  As a final remark, it is worth mentioning that the construction of the Galileon action involving simultaneously a scalar and a vector field is currently underway \cite{alejandro}.

\ack  Y.R. wishes to thank E.~Allys, P.~Peter, and J. P. Beltrán Almeida for the nice collaboration they were involved that allowed him to immerse in the beautiful world of the Galileon/Horndeski theories.  He also wishes to thank Ryo Namba and Lavinia Heisenberg for interesting and useful recent correspondence.  We acknowledge Alejandro Guarnizo and Luis Gabriel Gómez for their help with the analysis of some terms in the generalized $SU(2)$ Proca action.  This work was supported by COLCIENCIAS - ECOS NORD grant
number RC 0899-2012 with the help of ICETEX, and by COLCIENCIAS grant numbers 110656933958 RC 0384-2013 and 123365843539 RC FP44842-081-2014.

\appendix

\section*{Appendix}

\setcounter{section}{1}

We will show in this appendix why the $\mathcal{L}^{\rm Gal,bis}_{4,A}$ Lagrangian, uncovered in Ref. \cite{Allys:2016jaq}, is redundant\footnote{This is an unpublished proof by Ryo Namba.  We thank him for sharing the proof and giving us permission to reproduce it here.}.

Let's start with the definitions
\begin{equation}
\mathcal{L}^{\rm Gal,bis}_{4,A} \equiv g_4(A^2) \ A^\mu \tilde{F}_{\mu \nu} S^{\nu \lambda} A_\lambda \,,
\end{equation}
and
\begin{eqnarray}
F_{\mu \nu} &\equiv& \partial_\mu A_\nu - \partial_\nu A_\mu \,, \\
S_{\mu \nu} &\equiv& \partial_\mu A_\nu + \partial_\nu A_\mu \,.
\end{eqnarray}
Thus,
\begin{eqnarray}
\mathcal{L}^{\rm Gal,bis}_{4,A} &\equiv& g_4(A^2) \ A^\mu \tilde{F}_{\mu \nu} S^{\nu \lambda} A_\lambda \nonumber \\
&=& g_4(A^2) \ A^\mu A_\lambda \tilde{F}_{\mu \nu} (F^{\lambda \nu} + 2 \ \partial^\nu A^\lambda) \nonumber \\
&=& g_4(A^2) \ A^\mu A_\lambda \left(\frac{1}{4} \tilde{F}_{\rho \sigma} F^{\rho \sigma} \delta^\lambda_\mu + 2 \tilde{F}_{\mu \nu} \  \partial^\nu A^\lambda \right) \nonumber \\
&=& g_4(A^2) \left(\frac{1}{4} A^2 F \cdot \tilde{F} + A^\mu \tilde{F}_{\mu \nu} \  \partial^\nu A^2 \right) \nonumber \\
&=& \frac{g_4(A^2)}{4} A^2 F \cdot \tilde{F} + g_4(A^2) \partial_\nu A^2 A_\mu \tilde{F}^{\mu \nu} \,,
\end{eqnarray}
where we have used, in the third line, the identity $\tilde{F}_{\mu \nu} F^{\lambda \nu} = \frac{1}{4} \tilde{F}_{\rho \sigma} F^{\rho \sigma} \delta^\lambda_\mu$ derived for the first time in Ref. \cite{Fleury:2014qfa} and employed in Ref. \cite{Allys:2016jaq}.

Let's now define
\begin{equation}
G_4(A^2) \equiv \int^{A^2} d(A^2)' g_4[(A^2)'] \,,
\end{equation}
so that
\begin{equation}
\partial_\nu G_4(A^2) = G_{4,A^2} \partial_\nu A^2 = g_4(A^2) \partial_\nu A^2 \,.
\end{equation}
Therefore,
\begin{eqnarray}
\mathcal{L}^{\rm Gal,bis}_{4,A} &\equiv& \frac{g_4(A^2)}{4} A^2 F \cdot \tilde{F} + \partial_\nu G_4(A^2) A_\mu \tilde{F}^{\mu \nu} \nonumber \\
&=& \frac{g_4(A^2)}{4} A^2 F \cdot \tilde{F} + \partial_\nu \left( G_4(A^2) A_\mu \tilde{F}^{\mu \nu} \right) - G_4(A^2) \partial_\nu (A_\mu \tilde{F}^{\mu \nu}) \nonumber \\
&=& \frac{g_4(A^2)}{4} A^2 F \cdot \tilde{F} + \partial_\nu \left( G_4(A^2) A_\mu \tilde{F}^{\mu \nu} \right) - G_4(A^2) \left(\frac{1}{2} F_{\nu \mu} \tilde{F}^{\mu \nu} + A_\mu \partial_\nu \tilde{F}^{\mu \nu} \right) \,. \nonumber \\
&&
\end{eqnarray}
The Bianchi identity $\partial_\alpha F_{\beta \gamma} + \partial_\gamma F_{\alpha \beta} + \partial_\beta F_{\gamma \alpha} = 0$ leads us immediately to conclude that $\tilde{F}^{\mu \nu}$ is divergenceless.  The final conclusion is, therefore,
\begin{equation}
\mathcal{L}^{\rm Gal,bis}_{4,A} \equiv \frac{g_4(A^2) A^2 + 2G_4(A^2)}{4} F \cdot \tilde{F} + \partial_\nu \left( G_4(A^2) A_\mu \tilde{F}^{\mu \nu} \right) \,,
\end{equation}
where the first term belongs to $\mathcal{L}^{\rm Gal}_{2,A}$ and the second one may be discarded as it is a total derivative.  The curved spacetime generalization of this result still applies due to the Bianchi identity of the Riemann tensor that guarantees that $\tilde{F}^{\mu \nu}$ is still divergenceless.

\section*{References}
\bibliography{bibli}

\end{document}